\documentclass[twoside,11pt]{article}

\usepackage{obs_study_style}

\usepackage{moreverb,url}

\usepackage[colorlinks,bookmarksopen,bookmarksnumbered,citecolor=red,urlcolor=red]{hyperref}

\newcommand\BibTeX{{\rmfamily B\kern-.05em \textsc{i\kern-.025em b}\kern-.08em
T\kern-.1667em\lower.7ex\hbox{E}\kern-.125emX}}

\heading{}{}{}{}{}{Jordan Rodu and Michael Baiocchi}

\ShortHeadings{When black box algorithms are (not) appropriate}{Rodu and Baiocchi}
\firstpageno{1}

\begin{document}

\title{When black box algorithms are (not) appropriate}

\author{\name Jordan Rodu \email jordan.rodu@virginia.edu \\
       \addr Department of Statistics\\
       University of Virginia\\
       Charlottesville, VA 22904, USA 
       \AND
       \name Michael Baiocchi \email baiocchi@stanford.edu \\
       \addr Epidemiology and Population Health\\
       Stanford University \\
       Stanford, CA 94305, USA}
\maketitle

\begin{abstract}%

In the 1980s a new, extraordinarily productive way of reasoning about algorithms emerged. In this paper, we introduce the term ``outcome reasoning'' to refer to this form of reasoning. Though outcome reasoning has come to dominate areas of data science, it has been under-discussed and its impact under-appreciated. For example, outcome reasoning is the primary way we reason about whether ``black box'' algorithms are performing well. In this paper we analyze outcome reasoning's most common form (i.e., as ``the common task framework'') and its limitations. We discuss why a large class of prediction-problems are inappropriate for outcome reasoning. As an example, we find the common task framework does not provide a foundation for the deployment of an algorithm in a real world situation. Building off of its core features, we identify a class of problems where this new form of reasoning can be used in deployment. We purposefully develop a novel framework so both technical and non-technical people can discuss and identify key features of their prediction problem and whether or not it is suitable for outcome reasoning.

\end{abstract}


\section{Introduction}
\label{sec:intro}
It is exciting to witness the development of flexible, fast, and useful predictive algorithms. Algorithms are driving cars, identifying breast cancers, enabling globe-spanning businesses, and organizing unprecedented amounts of information; prediction provides a strong foundation for technological innovation and scientific breakthroughs. The excitement about these algorithms is warranted; these achievements are unparalleled in history. A very natural question for members of the business, government, and academic communities is: ``how can we use them?'' This simple question is more complicated than it first appears because modern predictive algorithms have several very distinctive features, including: some of the most successful algorithms are so complex that no person can describe the mathematical features of the algorithm that gives rise to the high performance (a.k.a. ``black box algorithms''). There is an important tension right now because these extraordinary black-box algorithms exist -- with so much potential to do good -- despite deep uncertainty about when and how to use them. And this tension is warranted: for all of their achievements, black-box algorithms have shown to be unpredictably brittle in the real world, failing without much warning.  This is a consequence of how they are developed. 

These algorithms have come into existence through a confluence of innovations. Some of these innovations are practical (e.g., the price of computing has continued to drop), some are market-based (e.g., online platforms have proved to be profitable business models and so corporations have funded much of this research and development), some are due to political decisions (e.g., emphasis on funding education and development in STEM fields has created a large pipeline of data scientists), but a major -- yet also under-appreciated -- shift has come from a \textit{new way to reason using data}, a form of reasoning that does not require slow-moving mathematical proofs. While understanding black box algorithms in general is not possible, by understanding how they are being developed and assessed we can understand what situations are more -- and less -- compatible or safe for their use. To provide guidance on using prediction algorithms, this paper offers a new framework for stakeholders (e.g., business people, government officials, non-statistically minded academics, individuals affected by predictions) to discuss and critique the use of these algorithms. For reasons which will become clear, we call this framework MARA(s).

\subsection{Goals of this paper}

This paper addresses several important concerns about using machine learning, or black box algorithms, in real-world scenarios.
\begin{enumerate}
    \item The framework introduced in this paper allows for a principled way to determine if a black box algorithm is appropriate for a prediction problem (see sections \ref{sec:certification_intro} and \ref{sec:certify}).
    \item This paper provides a framework for discussing and debating the suitability of black box algorithms that can accommodate both technical and non-technical individuals, without relying on the translation or interpretation of technical matters for non-technical individuals (see sections \ref{sec:certification_intro} and \ref{sec:certify}).
    \item The framework provides guidance for easily generating powerful, novel critiques of deploying black box algorithms in particular situations (see, for instance, subsection \ref{subsec:generating_critiques} for an example in the context of recidivism algorithms).
    \item The framework aids users in identifying ways that they can better modify and design their approach to a particular problem of interest so that black box algorithms can be used in targeted ways (see subsection \ref{causal_preds}).
\end{enumerate}

In the next section we describe the common task framework (CTF), which is the intellectual-engine that has driven much of the innovation in machine learning.  Through a careful assessment of the CTF, we identify a novel form of reasoning utilized in the development of machine learning algorithms, which we call ``outcome reasoning.''  We argue that, instead of debating about the appropriateness of deploying ``machine learning'' or ``black box algorithms'' in different scenarios, we should consider the suitability of outcome reasoning for the particular situation.  The rest of the paper introduces MARA(s), a framework that offers guidance for how to assess the fitness of outcome reasoning for a problem at hand, and improve this form of reasoning to support algorithms during deployment in the real world.

\section{Background: the Common Task Framework}
Below we give a short introduction to the intellectual-engine that has driven much of the recent innovation in machine learning, and the one that has consequences on how algorithms are deployed.  The framework offered in this paper is motivated by a careful study of the properties of the common task framework.

The Common Task Framework (CTF) \citep{Liberman2010-gt, Donoho2017-wx, Breiman2001-ft} provides a fast, low-barriers-to-entry means for researchers to settle debates about the relative utility of competing algorithms.  This is in contrast to the traditional use of mathematical descriptions of the behavior of an algorithm, or simulations of the algorithm's ability to recover parameters of a data generating function. Many readers are likely familiar with the CTF even if the name is unfamiliar; the NetFlix Prize \citep{Bennett2007-em} and Kaggle competitions are excellent examples of this framework. The key features of the CTF are: (a) curated data that have been placed in a repository; (b) static data (all analysts have access to the same data); (c) a well defined task (e.g., predict y given a vector of inputs x for previously unobserved units of observation); (d) consensus on the evaluation metric (e.g., the mean squared error of the predictions from the algorithm on a set of observations); and (e) an evaluation data set with observations which have not been accessible to the analysts. Today, in practice, some of the features of the CTF are relaxed. In particular, outside of the major competitions, feature ``e'' is often self-policed - i.e., the analyst has direct access to the evaluation data set. When performed correctly, the CTF gives us a way of justifying a claim that ``Algorithm A performs better than Algorithm B on these data sets.''

More than just a form of justification, the CTF provides an efficient environment for development. The data exist already. All analysts have access to these common data so many people can work on the problem at the same time. Fast computation takes the place of proving theorems, and performance is quickly assessed using held-out data. The consequences of a poorly performing prediction algorithm in the CTF are minimal –- e.g., after a failure the analyst tweaks the algorithm and tries again. Fundamentally, the CTF takes complex real-world problems and sand-boxes them.  

In the CTF, because there is a specific performance metric, there is less ambiguity in the relative ordering of the algorithms conditional on a particular dataset. The ordering gives rise to a ranking of the algorithms, and the public display of these rankings are called `leader boards.'' Leader boards are cited as using competition to motivate analysts to reach high levels of performance \citep{meyer2011verification, costello2013seeking, boutros2014toward}. The underlying logic of using leader boards is new and productive. The CTF/leader boards are the most prominent example of the form of reasoning we call ``outcome-reasoning.''  Given the features described in the previous two paragraphs, outcome reasoning -- if appropriate -- is preferred. However, many people are deploying black-box algorithms, which rely on outcome-reasoning, in problem settings when outcome-reasoning is unavailable.  In these problem settings, ``model-reasoning'' should be used). Many current debates in the literatures about the suitability of black box models hinge on (mis)understandings about what kind of reasoning is appropriate for given problems  (the contrast between forms of reasoning is discussed in detail in \ref{subsec:intuition}. One side frames the challenges in terms that are specified by model-reasoning; the other side responds with adaptations to the CTF which do not map onto the points that are most contention in traditional model-reasoning (e.g., there is no need for a ``data generating function'' in outcome-reasoning). We suspect that those most familiar with model-reasoning may feel better prepared to discuss and critique algorithms generated using outcome-reasoning if its key features are better described. The goal of this paper is not to resolve these debates between the two camps but to provide a useful framework for understanding the type of problem amendable to outcome-reasoning.

\section{Thinking in Terms of ``Outcome-Reasoning'' Instead of ``Black Box Algorithms''}
\label{sec:perspective_black_box}

\label{subsec:blackbox_type}

In this section we have a more detailed discussion of what we mean by ``black box'' algorithms. We also point out how people are convincing themselves that black box algorithms ``are working.'' 

The classification laid out in \cite{burrell2016machine} is quite useful for discussing how an algorithm becomes opaque: [Type 1:] opacity as intentional corporate or state secrecy, [Type 2:] opacity as technical illiteracy, and [Type 3:] an opacity that arises from the characteristics of machine learning algorithms and the scale required to apply them usefully. The root-causes of the opacity are interesting, and have implications for how to remove the opacity. Important research focuses on how to reduce the opacity of the algorithms without losing the strengths offered by these algorithms. Instead of focusing on the causes of the opacity, we focus on how one goes about convincing others that the black box algorithm is suitable for deployment. 

For the sake of argument, consider an algorithm that satisfies all three definitions of black boxes as offered by Burrell. Taking this as our example, what do we mean when we say this algorithm ``works well''? Few understand how this complex algorithm is implemented, and none argue the behavior of the algorithm is well-understood. So describing how it links variation in the input space to variation in the outcome space is not an option; we cannot use model-reasoning. It appears that outcome-reasoning is the dominant way to reason about the performance of this kind of algorithm. People believe this complex algorithm works not because they believe the algorithm \textit{should} work for some problem because of how the algorithm functions, but rather because people have observed these algorithms working -- in a CTF-sense -- by out performing other algorithms on predictions tasks. If these algorithms are not justified by outcome-reasoning then what argument, based on data, is used to justify these algorithms? 

While there are three distinct causes of black boxes, recognize that regardless of type we justify a black box by using outcome-reasoning. While the CTF was first developed to address the development of Type 3 black boxes, the logic of outcome-reasoning has been co-opted to justify the other types of black boxes. With outcome-reasoning, organizations can try to convince stakeholders to use their algorithm while withholding details of how their Type 1 black box works. Without outcome-reasoning it is possible fewer Type 1 black boxes would be deployed, because it would be harder for corporations and state actors to convince stakeholders to accept the algorithm's utility without allowing a deeper interrogation of their algorithm (i.e., blocking any chance for model-reasoning). It is interesting to think about how outcome-reasoning has facilitated the proliferation of Type 2 black boxes -- e.g., by lowering the barriers to assessing an algorithm and therefore lulling more stakeholders into feeling comfortable deploying these algorithms. 

Outcome-reasoning does not require as detailed engagement with the behavior of an algorithm, so more people can reason about the relative performance of algorithms than if they were required to use model-reasoning. It is possible that the simplicity of outcome-reasoning has led to overconfidence from non-technical stakeholders as outcome-reasoning feels more accessible than model-reasoning. The work on algorithmic fairness has helped draw attention to the importance, and subtlety that data set selection and performance metric selection play in outcome-reasoning.

\section{Introduction to MARA(s)}
\label{sec:certification_intro}

In this section we introduce a framework, derived from the CTF, for evaluating if a prediction problem is suitable for using outcome-reasoning. Here we describe the MARA(s) framework in the simplest possible way so that we can quickly introduce some basic, instructive examples that will help the reader understand outcome reasoning and the MARA(s) framework (see \ref{sec:cases}).  We then return to a more detailed description of the MARA(s) framework that benefits from exposure to the simple examples (see \ref{subsec:taxonomy}).

In any interesting prediction problem, errors will occur. Outcome reasoning encourages a different view of how errors in prediction can and should be used to improve the algorithm. This view of how errors can be used is often off-putting to data scientists who largely use model-reasoning to assess algorithms. That unease is because outcome-reasoning targets \emph{different kinds of prediction problems} than model-reasoning. 

In model-reasoning, the data analyst works with stakeholders to propose a model that maps to what is known about the prediction problem and then the model is fit. In a simplified telling of how model-reasoning works: if (i) the errors produced by the model match the modelling assumptions and (ii) the parameters implied by the fit of the model are not too divergent from what we believe about the problem then we are comfortable deploying the model into the real world. Note that we have two checks before deployment: [technical] are modeling assumptions violated? and [belief-based] is the model fit compatible with what believe to be true. By examining these two aspects of the model fit, we can get buy-in from stakeholders before deploying the algorithm into the real world. Thus when the model starts to accumulate errors in the real-world we have already achieved some level of buy-in. But in outcome-reasoning we do not have these checks, so we need a different way of getting buy-in from stakeholders before deploying. In this section we propose a framework for getting this buy-in that we call MARA(s).

We classify problems using four features (``problem-features''), which we refer to collectively with the mnemonic ``MARA'': 
\begin{enumerate}
    \item \textbf{[measurement]} ability to measure a function of individual predictions and actual outcomes on future data,
    \item \textbf{[adaptability]} ability to adapt the algorithm on a useful timescale,
    \item \textbf{[resilience]} tolerance for accumulated error in predictions, and
    \item \textbf{[agnosis]} tolerance for potential incompatibility with stakeholder beliefs. 
\end{enumerate}

 Primary stakeholders classify their problem as either ``satisfying" or ``not satisfying" each problem-feature individually.  In some settings it may be more appropriate to relax the binary classification, which is discussed in detail the ``moving from binary features to continuous features'' subsection. If a problem satisfies the MARA problem-features then the problem is suitable for outcome-reasoning. If the problem fails to satisfy even one of the features then the problem requires a more complex form of reasoning to justify the algorithm's deployment -- i.e., model-reasoning (discussed in detail in the ``model- and outcome-reasoning'' subsection).

We arrive at MARA(s) by examining the features of the CTF and modifying those features to be appropriate to ``live" problems, providing a principled foundation for assessing the generalizability and transportability of an algorithm into the real-world. That is, if an algorithm was developed using the CTF then MARA(s) provides a way of identifying real-world prediction problems that resemble a CTF setting. MARA(s) provides language to clarify a problem's features and facilitate debate among stakeholders about the suitability of using outcome-reasoning to evaluate the performance of an algorithm.

 We also emphasize that MARA(s) explicitly starts with the stakeholders. Starting with stakeholders has two major implications: (a) the focus of this assessment is based on understanding the problem itself rather than the algorithm, and (b) a different set of stakeholders can yield a very different conclusion about the appropriateness of a black box algorithm (see the recidivism example below for how this works). This emphasis on stakeholders has more in common with model-reasoning --where one of the first steps is to ``describe a model''-- and less with Leo Breiman's take on algorithmic culture-- that data analysts concede the complexity of the data generating function and instead think about the properties of the algorithm \cite{Breiman2001-ft}. Breiman's line of thinking has led to excising stakeholders (e.g., experts, laypeople) from the process, and an increased emphasis on the technical aspects of the algorithm and its performance.

We flag an important issue here: there can be deep ethical concerns about how stakeholders are included and excluded. We do not engage these concerns in this paper, but we do want to emphasize that we offer a language for critique: (i) ``I should have been a stakeholder when the problem was being defined.'' (ii) ``If I had been a stakeholder then I would have argued that this problem fails both adaptability and agnosis.'' It is an important feature of the MARA(s) framework to provide language that can express that different sets of stakeholders -- e.g., $s_1$ and $s_2$ -- will have different assessments of satisfying the MARA: $MARA(s_1) \neq MARA(s_2)$.  A different theory is needed to think about how we should go about including and excluding stakeholders.  We provide further discussion about stakeholders in the supplementary material.

\begin{figure}
    \centering
    \includegraphics[width=1\linewidth]{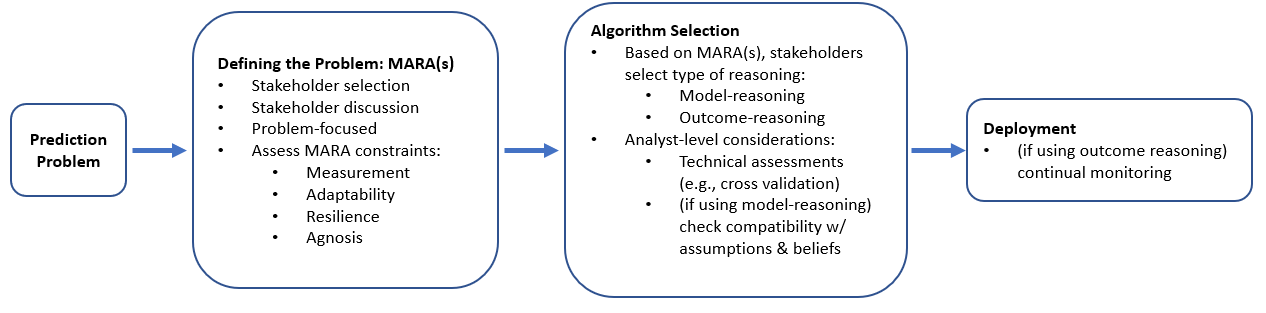}
    \caption{Workflow}
    \label{fig:workflow}
\end{figure}

This framework implies a workflow (see Fig \ref{fig:workflow}) in which primary stakeholders first engage a prediction problem through MARA(s).  This occurs prior to any technical considerations (e.g., using cross validation to assess the fit of the algorithm).  Only after classification under MARA(s), analysts can identify suitable algorithms that adhere to the proper form of algorithmic reasoning, and can assess the potential algorithms for their technical merits.  Finally the analyst can deploy the algorithm with proper re-assessment, depending again on the class of reasoning chosen through MARA(s).

\section{Simple examples of each part of MARA}
\label{sec:cases}

In this section we discuss several common prediction problems. Each example was chosen to highlight particular aspects of MARA(s). Each of the MARA problem-features is considered, as well as how stakeholder selection impacts these considerations. 

The goal of this section is to provide easy to follow examples of each problem-feature, not to provide a detailed working through of an actual prediction-problem. To see MARA(s) used in an in-depth way through a real-world problem see section ``using MARA(s) to reason about a recidivism algorithm.''

\subsection{Canonical case: recommendation systems}
The goal of a recommendation system is to introduce users to products of interest.  Training data for a recommendation system often is represented as a sparse matrix with individuals on the rows and products on the columns.  If the $i^\mathrm{th}$ user has engaged with the $j^\mathrm{th}$ product and rated it, the $ij$ entry of the matrix will contain the user-product rating.  Other entries will be blank.  To select algorithms, some of the known entries are obscured, and the task is to predict the rating for those user-product pairs. In deployment, algorithmic performance can be measured as a function of the individual outcome of each recommendation, which might take the form of a) acceptance of recommendation with a subsequent high rating, b) acceptance with a low rating, c) acknowledgement of the recommendation without taking it, or d) no acknowledgement of the recommendation.  Technically, only some of the predictions are verified in the wild.  When the system predicts that the user will assign a low rating to a product, that product will not be recommended, and that rating might not be verified.  But the ultimate goal of the recommendation system is to provide good recommendations to make money.  Often there is a large pool of products, many of which could be recommended to the user with great success.  The loss function, then, is asymmetric.  \textit{If} a product is recommended to a user, it is desirable that the product will obtain a high rating.  On the other hand, if a product that would otherwise have been enjoyed by a user is not recommended, because of the large pool of potential products, missing this product is not such a big deal.  In general, recommendation systems satisfy the MARA problem-features, though specific examples could be constructed in which some of the problem-features might fail to be satisfied.

\subsection{Varying stakeholders changes resilience and agnosis: financial trading}
Consider a high-frequency trading group with large reserves and a goal of profit maximizing. In order to construct a risk-diversified portfolio, one sub-goal in high frequency trading is to predict the instantaneous covariance between stocks.  While one cannot directly assess the accuracy of the covariance prediction, one could use an external measure of performance as a direct consequence of the prediction to monitor the success of the algorithm. For example, the group can monitor if the ongoing use of the algorithm increases the value of their holdings. In this setting, all four problem-features are satisfied.

In contrast, a manager of family wealth may require an algorithm that can be evaluated by the family so that they can assess whether or not the algorithm's anticipated behavior matches their beliefs about the market or to verify that the trading algorithm comports with their ethical concerns.  In this case, agnosis would not be satisfied. Similarly, a family may not be as resilient to losses as a large high-frequency trading group so may not satisfy the resilience requirement; that is, they don't have deep enough pockets to sustain financial losses as the algorithms are used and improved.

\subsection{Failing measurement: prediction in lieu of measurement}\label{sub_measurement}
An interesting application of prediction algorithms is to use them as cheap measurements in lieu of obtaining expensive, gold-standard labels.  Consider an automated system for triaging mild health symptoms.  Instead of using a telephonic nurse-based system, automated prediction algorithms could be used to offer some level of diagnosis and either recommend a patient seeks further help or not.  From the perspective of a health care administrator, triaging mild health symptoms may satisfy all four problem-features because the administrator considers outcomes of many patients who interact with the health system. But if administrators take as their goal the delivery of care to a particular patient then the problem fails measurement because the prediction is explicitly deployed in lieu of an actual diagnosis. That is, the point of such a prediction algorithm is to skip the burden of obtaining the desired measurement. 

Importantly, if the patient is included as a stakeholder, then the adaptability problem-feature is also violated since this is a one-shot prediction for this patient.  This is a general principle: problem-feature satisfaction depends on who is considered as a stakeholder. That is, in general, $MARA(s_i) \neq MARA(s_j)$.

\subsection{Failing agnosis and measurement: recidivism}\label{sub_recidivism}
In an algorithm is used to predict recidivism, if all defendants with a score above a certain threshold are incarcerated then we are unable to observe the correctness of our predictions for people who are incarcerated.  This arises from a missingness in the outcome, and will happen in general when the prediction algorithm causes changes in the outcome. The recidivism problem also violates the resilience and agnosis problem-features.  In the resilience problem-feature, the debate hinges on the stakeholders' concerns about depriving rights through unnecessary incarceration, balanced against possible future criminal acts.  Failure to satisfy agnosis stems from two concerns.  First, it is necessary to explain to the defendant why the decision to incarcerate was made.  Second, even if the algorithm were a flawless predictor, if it did so through morally repugnant means then stakeholders would need to know that it is achieving its predictions this way and these means would need to be debated by stakeholders (these kinds of concerns are often referred to as ``deontological'' -- roughly: having to do with ethical considerations about the actions taken rather than just concerns about the outcomes achieved). 

\subsection{Agnosis and causality: optimizing causal predictions}\label{causal_preds}

Suppose we run a website that can place only one of three ads -- corresponding to one of three items for purchase -- for each customer who arrives to the website. We are unsure of which ad to place for a given customer. Let us consider if a black box algorithm is suitable for use in learning an optimal assignment of ads.

In this setting, a useful algorithm estimates how much the probability of buying a product changes if the website shows a certain ad to a particular customer. At first pass it may seem best to assign customers to the ad that will increase their probability of purchase the most -- but that quantity is estimable only if the analyst knows how the probabilities of purchasing change given an ad, up to some tolerance. The prediction problem is thus to learn these probabilities and the question is whether or not this is a suitable problem for outcome reasoning.

Similar to the recidivism example above, for any particular person we can only measure the outcome for one of the ad placements. It is possible, though, to target a useful function of these outcomes: the average outcome after exposure to the ad placement conditional on some set of covariates (this is often referred to as the conditional average treatment effect and can often be used to build up other popular estimands). This can be done by having the algorithm assign the ad placements in such a way that for any person there is some chance of seeing each ad - i.e., $0 < p_1, p_2,p_3 < 1$. This is how randomized trials obtain causal estimates. In contrast to a uniform randomized controlled trial -- where the assignment to treatment is proportional to rolling a die with three equally weighted outcomes -- in this setting we can use some more sophisticated algorithm that may be able to obtain better estimates of the probabilities more efficiently. This kind of thinking has led to interesting work on contextual bandit theory, and other forms of adaptive trials. Thus, in some prediction settings with causal components, measurement can be achieved through introduction of a randomization. Yet there is something different between this problem and most randomized controlled trials (RCTs).

RCTs take as their goal to evaluate mechanisms for the observed change (for example: does increasing temperature lead to more plankton in this niche). The intent in RCTs is to link variation in the input space with variation in the outcome. Traditional RCTs fail agnosis because the researchers take as their goal generating information which can be assessed for its compatibility with existing knowledge and belief. In contrast in the setting of ad placements, the goal is to sell more product and there is no intent to understand the mechanisms giving rise to that increase. It may not bother the website deploying the ads if the black box being used to optimize the ad placement does it in a way that defies their beliefs. 

When humans are the subjects in RCTs, issues related to resilience are considered by the institutional review board. Less obvious, but still a challenge to the agnosis problem-feature is that an IRB may want to evaluate the assignment mechanism to assess whether it is assigning study participants in an ethical manner. If an RCT is using a black box to assign patients then this precludes oversight. 

When the MARA problem-features are satisfied, this gives rise to a peculiar problem type: an atheoretic randomized controlled trial with treatment assignment determined by a black box algorithm.

\section{Reasoning about algorithms}
\label{sec:certify}

In this section we compare two types of reasoning used for assessing an algorithm.  

\subsection{Model- and Outcome-reasoning}
\label{subsec:intuition}

In algorithmic prediction, there are two common ways of using data to reason about the performance of an algorithm. The first we discuss is model-reasoning, which is the more traditional form of reasoning exemplified by linear models. The second form of is outcome-reasoning.

Model-reasoning requires checking that the model conforms to current beliefs. For example consider a linear regression; we can use model-reasoning by verifying the direction of individual coefficients matches what is expected from our domain knowledge.  We can think of these checks as a mapping from the model, or ``parameters'' of the model, to the space of current beliefs.  In model-reasoning, it is therefore possible to hypothesize how a particular instantiation of a model, $\hat{f}$, will perform on future data without reference to the data used to fit the algorithm.  This provides solid ground on which experts in a field can debate and discuss the fitted algorithm and its suitability for future predictions in a concise manner, with discussions stemming from beliefs, and not potential difficult-to-find shortcomings of the data set or algorithm.  

In contrast, outcome-reasoning relies very little on beliefs, which primarily enter into consideration through choice of the performance metric.  In MARA(s), outcome-reasoning is extended to the ongoing, out-of-sample prediction setting. If the four problem-features are satisfied then the analyst and stakeholders can monitor the performance metric during the deployment phase in order to assess whether or not their algorithm is working.

The two types of reasoning lead to two different kinds of thinking when comparing algorithms. Model-reasoning tends to involve discussions of parameters and the algorithm's ability to faithfully recover the parameters. That is, model-reasoning forces the analyst to think carefully about how changes in the covariates should be linked to variation in the outcome (e.g., should we predict that a taller person weights more than a shorter person?). But when divergent algorithms are compared -- e.g., say an ARMA(p,q) is compared to a decision tree -- it is quite challenging to translate between different conceptualizations of how the input space is linked to the outcome space. Consequently, given the challenge of translating between algorithms using model-reasoning, comparisons tend to be pairwise and slow. In stark contrast, outcome-reasoning assiduously avoids any debates in the input space. Instead, outcome-reasoning operates in the space of the outcome -- where all candidate algorithms must operate. An analogy to capture this dynamic: consider two economies. Model-reasoning is a bit like a barter-based economy; each transaction requires careful consideration of idiosyncratic features and how much the parties need each of the products. An economy that uses currency to store value allows a lower friction form of transactions; each product's value is translated into the currency and then comparisons can be made rapidly between different products. Now imagine these two economies and their ability to develop, innovate, and scale.

Many, though not all, concerns about black box algorithms can be framed as issues of extrapolation. While the CTF offers a foundation for comparing algorithms' performance on currently available data, the CTF alone does not offer a principled foundation for reasoning about the future, out-of-sample performance of an algorithm.  Without access to model-reasoning, the mechanisms for reasoning through performance on future data, and not just held-out data, are limited. Such reasoning would require careful consideration of the interaction between properties of the data set and the properties of the algorithm.  In a setting that uses a black box algorithm that requires massive training data, understanding the data itself can be impossible.  Even under a smaller data regime, the task of unpacking the data/algorithm interaction can be difficult, if not impossible.  Indeed one common fix when a black box fails is to add data to the training data set in hopes that a new fit on the new data might remedy the failure.  The underlying cause for the failure with respect to the current $\hat{f}$ often remains unknown.

An important distinction between the two modes of reasoning: model-reasoning allows for detailed debate to happen before the deployment of the algorithm, whereas outcome-reasoning affords assessment purely post-deployment. If an algorithm ``beat out'' all other contender algorithms during a CTF/leader board competition then there is no guarantee that it will continue to perform well on future data, because the CTF/outcome-reasoning does not provide a foundation for assessing the algorithms' performance on data not currently available. Outcome-reasoning requires the algorithms to generate predictions before they can be evaluated.

\subsection{The MARA framework}
\label{subsec:taxonomy}

We now discuss the four problem-features in more detail. Collectively, we refer to the four problem-features as MARA.

\subsubsection{Problem-feature 1: measurement}
The first problem-feature is the ability to measure a function of the predictions and actual outcomes on future data.  Let $y^*$ be the value of a future outcome associated with predictors $x^*$, and $\hat{f}(x)$ denote the estimated prediction function.  For some agreed-upon notion of close, this problem-feature describes the ability to track whether $\hat{f}(x^*)$ is close to $y^*$, measured as $g(y^*, \hat{f}(x^*))$, within some reasonable tolerance.  

This is the most foundational problem-feature for the MARA(s) framework. If this problem-feature isn't satisfied the analyst will not be able to verify if the algorithm is performing well. The use of a black box  model for a problem that doesn't satisfy measurement requires faith. If this problem-feature is satisfied then the algorithm's performance can be monitored after deployment by monitoring the error function $g()$ (notably, both Google \citep{google_undated-ux} and Uber \citep{Hermann2018-yj} include monitoring predictions after algorithm deployment as a critical component of their machine learning workflows). Without this feedback mechanism, assessment must happen before deployment. In this case, our framework requires that the analyst pursue model-reasoning.

\subsubsection{Problem-feature 2: adaptability}
Problem-feature 2 is the ability to adapt the algorithm on a useful timescale.  In some settings, upon discovering errors in prediction, an algorithm can be updated quickly and will be presented with sufficient opportunity to update.  In other settings, the underlying dynamics of the population of interest change at a rate such that those changes dominate algorithm adaptations from observed error.  The latter situation renders predictions as one-shot extrapolations, at which point the observation of the function $g()$ is useless.  

For instance, predicting the outcome of the United States presidential election depends on measuring the ebb and flow of priorities of the voting population.  With an algorithm assessed under outcome-reasoning, the lessons learned from prediction errors in one election may not be informative for the next election because the underling priorities of the population may have shifted. Another common violation of adaptability is when the deployment of the algorithm itself changes the way the outcomes are generated; this phenomenon has been described many times in policy settings - the Lucas Critique, Goodhart's law, and Campbell's law being famous formulations.

\subsubsection{Problem-feature 3: resilience}
Problem-feature 3 describes the stakeholders' tolerance for accumulated error in predictions. As errors accumulate, someone or some group will be held accountable. Some stakeholders will see errors in prediction as so intolerable as to bar any unjustified use of an algorithm, for example when an error in prediction may lead to a death or a false incarceration. On the other end, settings like recommendation algorithms may be viewed as having minimal consequences to errors in prediction. Most scenarios will be somewhere in between, where the stakeholders are willing to trade off some unaccounted error in predictions against the accumulation of value gained from better predictions. If the group deploying the algorithm has large reserves relative to the accumulation of costs due to the accumulation of errors then the problem at hand satisfies resilience.

\subsubsection{Problem-feature 4: agnosis}
Problem-feature 4 describes tolerance for incompatibility with stakeholder beliefs. Stakeholders will hold beliefs about the process being predicted. Typically these will be expressed as how variation in the input space is linked to variation in the outcome, though other beliefs can also be strongly held (e.g., the outcome is bounded by 0 and 100). 

These beliefs may take the form of prior knowledge or scientific evidence (e.g., taller people tend to eat more calories, so an algorithm that lowers predictions for taller people may appear to be dubious).  Other beliefs may arise from moral or ethical concerns (e.g., racial information should not be used to assign credit scores).  In some cases, stakeholders' beliefs may be quite strong and thus they may not agree to deploying an algorithm to permits violations of their beliefs. 

The agnosis problem-feature requires both eliciting and clarifying the stakeholders' beliefs about the problem. It also requires understanding stakeholders' comfort with the algorithm violating their beliefs.

\subsection{Problem-feature discussion}

In the ``examples'' section we took great care to isolate examples so that each problem-feature appeared as clear and distinct as possible. In reality, these features interact and in practice should be discussed collectively as MARA. We encourage the use of ``MARA(s)'' to emphasize the role of the stakeholders in assessing the four problem features.

Model-reasoning requires deductive reasoning, meaning that we understand the mathematical structures of the model well enough so that once decoupled from the data it was fit on, stakeholders can reason about future behavior of the algorithm.  Methods of assessing an algorithm that are inductive cannot be used for model-reasoning.  Inductive reasoning is contingent and depends on the data in hand (e.g. recycled predictions) or on details of hypothesized, future, out-of-sample data.  Methods of assessment built off of these are attempting to approximate model-reasoning.

\subsection{Moving from binary features to continuous features}
For simplicity we have presented the problem-features as binary, as if a prediction problem either satisfies a problem-feature or does not.  But in reality that is too restrictive.  We believe that in practice MARA(s) will be implemented with continuous features-- a prediction problem might not completely satisfy, nor fail to satisfy, a problem-feature.  We leave a full specification of continuous MARA(s) for future work-- work that we hope will engage diverse thinkers across many disciplines.  To help lay the foundation for that work, we provide an example extension of MARA(s) to continuous problem-features.

A Pew Research Center study \citep{lam_hughes_kessel_2020} looked at party and gender gaps in postings about sexual assault on Facebook by members of Congress.  The researchers needed to label $44,792$ posts as either having or not having content about sexual assault.  By design, this problem fails to satisfy problem-feature 1, as the whole point of using an algorithm is to avoid having to measure all of the posts.

In that study, the researchers hand-labeled about $500$ posts and trained a classifier using outcome-reasoning.  After using the classifier on the rest of the posts, the researchers spot-checked a small handful of posts to gauge the accuracy of the classifier on ``unseen data''.  Ultimately, they found that women posted more than men about sexual assault, and that democrats posted more than republicans.  A binary MARA(s) suggests that this was the wrong approach because of a failure of measurement.  But to most people the use of outcome-reasoning feels OK here.  Why?  

First, the model-reasoning approaches available to the researchers would not have been as powerful as the outcome-reasoning approaches.  This in itself would not be a compelling reason to use outcome-reasoning; just because one form of reasoning is problematic does not make another less problematic.  It does suggest a potential area in which model-reasoning approaches should be improved.  But in extreme cases (where lives-- or at least significant resources-- are at stake) human labeling, though expensive, remains a competitive alternative to outcome-reasoning if model-reasoning is not viable.

Second, the conclusions of the study, while interesting and important, are not surprising.  Had the conclusions suggested that men (or republicans) posted more about sexual assault, surely the researchers would have put considerable time into investigating why this was the case.  It would have been a surprising result, and further investigation would have either prompted a more nuanced understanding of why people were posting about sexual assault, or would have exposed idiosyncrasies in the dataset that led to misclassification errors.  Again, while comforting, this confirmation of our prior beliefs is not in itself justification for using outcome-reasoning.  Rather, it shows how outcome-reasoning could suggest further lines of inquiry if it does not conform to those beliefs.

Third, while we do learn from this study-- it adds evidence to the idea that there is a gap in the importance members of Congress assign the issue of sexual assault-- the consequences of potential errors are not severe. Nobody dies if the algorithm makes mistakes.  Important resources are not diverted from those in need.  Were this a study that could, for instance, re-direct millions of dollars of funding, potentially causing a negative impact to people in need, we would not feel as comfortable with an algorithm that leaves the classification of $44,000$ posts unchecked by human observers.

This third point is key.  Neither the relative power of outcome-reasoning vs model-reasoning approaches, nor the unsurprising conclusions of the study, could mitigate a severe consequence as a result of the study.  The third point speaks to resilience.  
We do care about the conclusions reached; this study is not simply an exercise.  But failure in the algorithm either leaves us with the same impressions we had before the posts were classified (in the case that the actual conclusions of the Pew Research Center study were wrong) or prompts us to further investigation (in the case that conclusions of the Pew Research Center went against our preconceived notions).

How does a continuous MARA better describe this kind of problem?  A continuous spectrum of resilience implies a knob that controls some aspect of measurement.  In the case laid out here, it controls how many posts we verify.  Because resilience to incorrect classification is high in this case, spot-checking a relatively small number of posts is reasonable.  As the consequences grow more severe, and hence our resilience diminishes, more checking is required.  In the extreme case, where we completely fail to satisfy the resilience problem-feature, checking all the posts is required under outcome-reasoning\footnote{Importantly, this does not imply that resilience is not of concern when we have full measurement.  In the case presented here we receive all unseen observations as a single batch.  In many cases, we may get observations sequentially.  By the time we have verified our prediction, that prediction has already been acted upon.  If we cannot tolerate acting on incorrect predictions, then resilience is low.}.

There are other ways in which we may fail measurement in a binary MARA(s), but can still take advantage of outcome-reasoning in the continuous version.  For instance, in some cases, we may not get to directly observe if our predictions are accurate, but can measure a proxy.  If our problem implies a situation where our resilience to error is low, we might require that proxy to be highly correlated with a hypothetical measurement of our success of prediction.  In problems where our resilience is high, we might be OK with a proxy that is less correlated.  Here, resilience does not toggle the number of outcomes we observe, as it does in the Pew Research case, but rather the quality of our measurement.  While MARA(s) does not provide a numerical threshold for choosing (for instance) the number of observed predictions, or the quality of our proxy, it provides a framework and language for debating-- among both technical and non-technical stakeholders-- the appropriateness of both outcome- and model-reasoning to the problem at hand.  In many scenarios, similar to utility functions from the econometrics literature, applied researchers working with stakeholders will find that a quantitative description (at least a good approximation) of the MARA(s) problem-features is available.

\section{Using MARA(s) to Reason About a Recidivism Algorithm}
\label{sec:MARArecidivism}

In this section, we work through an example and focus on how MARA makes reasoning about a prediction problem clearer. Consider an example where a particular county-court wants to have a decision support tool to quantify the potential for recidivism. (Note: it is not clear to us that a decision support tool in this setting is a wise decision. But this is a scenario that has occurred, and it can be framed as a prediction.) The Court announces its interest and requests proposals from companies to create a decision support tool. Several hundred companies bid on the contract.

Recognizing that this can be thought of as a prediction problem, the Court provides a data set to the companies and then holds a contest using the Common Task Framework. The Court can thus rank the performance of the algorithms using their desired metric(s). The three top performers are kept and move on to a new round of consideration. The results of the competition were as follows: Company A used a fantastically complex algorithm and out-performed the other algorithms in the contest. Company B, which performed a noticeably less successfully compared to Company A, used a proprietary algorithm that Company B believes should not be shared publicly (perhaps because they are concerned about people exploiting weakness of the algorithm). Company C, which placed third in the competition, uses an algorithm that is based on linear regression and they are willing to share it publicly. Because of its best-in-competition performance (and perhaps given other considerations such as speed and cost) the Court decides to move forward with Company A.

Given the MARA(s) framework, it should be recognizable that the data-driven part of the selection process above was based on outcome-reasoning, which may or may not be a good way to cut down the pool of competitors according to how stakeholders think about the MARA features. MARA(s) gives language to stakeholders (particularly non-technical stakeholders) to criticize the process as described above. The way we've told the story so far, it is not clear who selected the data set used for the competition and selected the prediction task(s) and performance metric. Even in using outcome-reasoning, there are important roles for the stakeholders. It is likely that if the stakeholders are (i) judges, prosecutors, politicians, and law enforcement officers then they will have different priorities than a group of stakeholders that includes (ii) public defenders, victims' families, and prisoner advocates. Again, even within an outcome reasoning situation there are vital roles for the stakeholders.

Potential stakeholders can offer more fundamental critiques of the process outlined above; they can reject outcome reasoning. While the contest demonstrated the ordering on the data in existence, the real question for the stakeholders is to reason about deploying the algorithm in future, unseen data. In order to proceed the stakeholders need to have a discussion of how they, as a group, believe the prediction problem they are interested in satisfies the MARA conditions. Given the sensitivities around depriving citizens of their rights, it is likely that both resiliency and agnosis are not satisfied for stakeholders like public defenders and prisoner advocates. Additionally, when an individual is incarcerated, it is not possible to assess the validity of the algorithm's prediction based on their outcome, so for anyone incarcerated there is no measurement available. 

Without recourse to MARA(s), given the relative prowess of the companies' prediction algorithms, it may feel hard for stakeholders to articulate their concerns and, further, to identify that outcome-reasoning should not be sufficient to convince them given the concerns they have with the prediction problem. Further, MARA(s) allows certain types of criticism to be clearer: (i) Company A's Type 3 black box is concerning because the stakeholders have strong beliefs such as deonotological concerns that no one can reason about. It is likely Company A needs to be disqualified. (ii) Company B's Type 1 black box may allow people within the company to reason about future functioning of the algorithm (if it is an algorithm that they can model-reason about) but the reasoning is not being done by the stakeholders and any assertions from Company B that the algorithm is consistent with the stakeholder beliefs requires a level of trust and possible verification (i.e., social-psychological arguments). It needs to be clear that Company B has not "successfully" articulated a data-driven argument that can convince stakeholders to the same level as Company C. Competing successfully in a CTF event does not warrant deployment.

It is important for stakeholders to know, and have the language to hold others accountable for, getting buy-in pre-deployment. This buy-in process centers on understanding if the MARA conditions are satisfied. If the MARA are not satisfied then stakeholders should request model-reasoning before deployment.  In the next section, we provide examples of using MARA(s) to generate fundamental, novel critiques to using black box algorithms for a recidivism algorithm.

\subsection{Example of generating critiques}\label{subsec:generating_critiques}

Smart folks are achieving good critiques of recidivism algorithms, but it has taken pioneering thinking and intuition. Here we provide an example of how important critiques can be overlooked if we do not train people to be aware of the assumptions underlying outcome reasoning.  Here we provide a handful of examples for how MARA(s) can be used to generate critiques.

Algorithms typically provide predictions of the probability of recidivism-- say $P(X_i)$ for individual $i$ -- and are used to inform sentencing (e.g., if $P(X_i) > C$ for some threshold $C$ then the defendant will be incarcerated). Critiques have largely focused on the differential impact of these predictions on subgroups (e.g., racial groups being treated differently by the recidivism model). The language in fairness focuses on the implied ethics of the algorithm -- e.g., perpetuating existing unethical practices of racism or sexism. Many of these debates have taken shape around fixing the target task -- e.g., creating better metrics to evaluate which algorithms are performing best. The arguments built around this line of thinking are so important, and so convincing, that they make many of us uncomfortable about the deployment of algorithms in the setting of predicting recidivism and their use in sentencing. These critiques may be all that are needed to convince someone to not deploy algorithms in this setting.  But they are not the only critiques available. They are not even the most devastating, statistically speaking.

Here, we re-frame the current debate on algorithms use to predict recidivism: because the algorithms used to create the predictions are ``black box'' in some sense (see the ``thinking in terms of 'outcome-reasoning' instead of 'black box reasoning''' section for discussion) an inspection of the algorithm cannot reveal how the algorithm will perform.  

Here are four questions -- derived from the MARA(s) framework -- that are statistically problematic for black box algorithms in this setting:

(i) After an algorithm has been fit and deployed for use by a court, how would we know if the algorithm makes an incorrect prediction for a particular defendant with $P(X)>C$? [Measurement]

(ii) What happens if major economic or political dynamics shift (e.g., pandemic, major recession); would the probabilities of recidivism change? How would the algorithm adapt? Could we train a new algorithm before the situation shifts again? [Adaptability]

(iii) For someone being ``judged'' by such an algorithm, what are the costs of it being correct vs incorrect in its prediction? Are those costs reasonable to the individual being judged? [Resilience]

(iv) In many criminal law settings, in addition to establishing the act was perpetrated by the defendant there is a need to establish the defendant's motivation -- which can be used to modify the punishment (e.g., first degree murder compared to negligent homicide). That is, in many legal settings, \emph{why} a set of actions were taken is material.  This is particularly true when there are major consequences to the defendant (e.g., depriving the defendant of rights). With this in mind, how is a black box algorithm (even one that has produced ``fair'' predictions on historical data) justified if the way it arrives at its conclusions are inexplicable? Said another way: if a defendant is to be deprived of their rights then are they owed more of a justification than ``this algorithm predicts a high probability of recidivism''? [Agnosis]

\section{Related Work}
\label{sec:algo}
Recently the CTF, while continuing to yield huge success in algorithmic development, has seen a host of criticisms.  One concern about the CTF is that datasets can be overfit over time \citep{Rahimi-winnercurse, Rogers2019-au, Van_Calster2019-yf, Ghosh2019-ho}.  Despite all attempts to protect against overfitting, idiosyncratic aspects of a particular dataset are learned when heuristic improvements yield improved predictions over the state of the art.  Additionally, and a potential corollary of this criticism, given two sets of reference data, it is often unclear why an algorithm performs well on one dataset, but poorly on the other.  This has led some to call for more resources to be dedicated to understanding the theoretical underpinnings of the algorithms that have achieved such huge success, hoping to avoid a catastrophic failure in the future.  There have been several debates on the relative merits of careful theoretical justification vs. rapid performance improvement (see \cite{Rahimi_undated-wz} and rebuttal in \cite{Lecun_undated-mq}; see also \cite{Barber2019-su}).  We do not enter this debate here.  However, the substance of the debate is important in MARA(s).  As can be seen in how these algorithms respond to different datasets, their performance is a complex interaction of the data, which can often be quite large, and difficult-to-uncover aspects of the algorithms.  While some in the aforementioned debates call for more theoretical understanding of these algorithms ahead of rapid innovation, we take a different tack and ask when we can deploy a black box algorithm through outcome-reasoning to make predictions in the wild.  For an algorithm whose future performance is justified using a measurement of the algorithm's success in the space of the outcome, as is done in the CTF, this framework recognizes an extension to the CTF, at very least satisfying the measurement problem-feature.  In problems that do not satisfy the measurement problem-feature, the algorithm is being used to extrapolate without $a priori$ justification, and we have no way of measuring -- or perhaps even being aware of -- failures.  By construction, such extrapolation does not exist in the static version of the CTF.

This paper also relates to debates in ethical machine learning through both the agnosis problem feature, as well as through stakeholder inclusion. This literature is new, rapidly expanding, and impactful; we suggest interested readers consult the following as solid entry points: \cite{mittelstadt2016, corbett2018measure, lum2016predict, kusner2017counterfactual, nabi2018fair, Wiens2019-zm}. We use the MARA(s) framework in the ``examples'' section to demonstrate how this framework can be used to clarify concerns of this nature -- see the examples on recidivism and prediction in lieu of measurement. 

In the public literature, most discussion of the CTF has been undertaken by David Donoho \citep{Donoho2017-wx, donohocomments}. (Note: it appears that much of the development of the CTF happened outside of the public-facing, academic literature.  See our correspondence with Mark Liberman in the supplementary files) Of particular interest, Donoho develops the notion of hypothetical reasoning in \cite{donohocomments}, exploring how analysts have developed ``models'' -- formalizations of their beliefs into statements of probability models -- to ``... genuinely allow us to go far beyond the surface appearance of data and, by so doing, augment our intelligence.'' Using language developed for MARA(s) we might say that satisfying the agnosis problem-feature means forgoing the advantages Donoho identifies accrue because of hypothetical reasoning. That may be a reasonable choice in some settings, but it should give pause to researchers interested in generating solid scientific evidence. Along similar lines, \cite{kitchin2014} gives a fascinating and early discussion of the epistemologies of the emerging big data revolution.

Finally, the MARA(s) framework is related to work on explainability/interpretability of algorithms. The conversations on these topics have been happening for many years, and in several distinct literatures, so understanding the foundational concerns and identifying the through lines of thinking can be challenging. We direct readers to two touchstone pieces in the literature as a good place to start: \cite{Breiman2001-ft} and \cite{shmueli2010explain, shmueli2011predictive}. Cynthia Rudin \citep{rudin2018stop} explores the suitability of two types of models, explainable machine learning models and interpretable machine learning models, in the context of high risk and low risk predictions. In Rudin's dichotomy, she warns against using explainable models in a high risk prediction due to our inability to make sense of the performance of the model despite the promise of explanation.  Instead, for high risk scenarios, she urges the practitioners to use interpretable models that can be linked directly to domain knowledge, and encourages researchers to put effort into finding suitable interpretable models where none exist.  In terms of our framework, Rudin is exploring the joint impact of the resilience and agnosis problem-features.  We direct the reader to Rudin's paper for details on why explainable machine learning models are not sufficient for what we identify as problems that require model-reasoning.

\section{Discussion}
\label{sec:disc}

The MARA(s) framework focuses on features of the prediction problem at hand, rather than the features of the algorithm.  The problem itself is selected by the stakeholders who have concerns that include accountability.  Understanding how stakeholders see the problem is the critical first step towards selecting an appropriate algorithm.  This framework directs attention to the four problem-features that stakeholders should assess: measurement, adaptability, resilience, and agnosis (``MARA''). Once assessed, the appropriate method for reasoning about the algorithm can be selected. 

In contrast to (but not in conflict with) MARA(s), there are other frameworks for decision-making about the suitability of an algorithm, technical in nature and useful for understanding the performance of different algorithms -- e.g. diagnostic tools or asymptotic performance -- but these are helpful after the method of reasoning has been selected.

While MARA(s) is a statement about the problem and not the algorithm, it does imply a loose structure to the set of possible algorithms.  One way to think about this implied structure is that model-reasoning methods are decoupled from the data, allowing for deductive reasoning about future performance, while outcome-reasoning relies on contingent, inductive reasoning.  Many of the current approaches to describing black-box algorithms are inductive in nature (see \cite{rudin2018stop}).  While these can be quite useful, they are still a qualitatively different form of reasoning.  This is a familiar distinction in the type of evidence we bring to problems, and the reader need not look hard to find examples in which deductive reasoning is a required component of our decision making. The gold-standard of inductive reasoning is randomized trials, but in the most consequential settings, the result of the most solid form of inductive reasoning does not provide sufficient justification.  For example, when approving a new drug, government agencies do not typically allow evidence from an atheoretical randomized trial to warrant approval.  Instead agencies require a detailed scientific hypothesis about how the drug's mechanism causes the outcome.  The addition of deductive reasoning and coherence across beliefs provides a firmer, evidence-based foundation. And yet, when appropriate, the use of outcome-reasoning is to be preferred because it is a powerful engine for producing the highest quality predictions. Outcome-reasoning is the intellectual-engine of modern prediction algorithms. 

If the data sets are interesting, the task is useful, and the performance metric describes an ordering that matches how the algorithm will be used then outcome-reasoning leads to an extraordinary consequence: it allows an analyst to bypass the slow, technical challenge of mathematically describing the behavior of the algorithm. Instead, outcome-reasoning allows the analyst to look at the joint distribution of predicted and observed outcomes and then rank performance of algorithms by creating statistical summaries. Outcome-reasoning leverages Tukey’s insight that \citep{tukey1986sunset}: ``In a world in which the price of calculation continues to decrease rapidly, but the price of theorem proving continues to hold steady or increase, elementary economics indicates that we ought to spend a larger and larger fraction of our time on calculation."

The power and popularity of the CTF has inspired extensions to prediction domains that are not traditionally investigated inside the framework.  For instance in \cite{wikle2017common} the authors propose an extension to spatial prediction which, among other additions, includes an abundance of relevant data sets of differing characteristics on which the algorithm must succeed, and additional metrics, like assessment of prediction coverage.  It seems reasonable that the CTF-SP will enhance rapid innovation of algorithms for certain types of problems, which is an exciting prospect for the spatial forecasting community.  However, we caution that, like all algorithms that are developed in the CTF, those algorithms that are not qualified to be reasoned about using model-reasoning and should only be used in a situation that permits outcome-reasoning.

The MARA(s) framework provides a language to help stakeholders and analysts communicate the key features of a problem and then guide the selection of an appropriate algorithm.  This language can also be used by algorithm developers to help identify areas for innovation.  For instance, in the ``examples'' section we discuss ``prediction in lieu of measurement'' and we are unaware of effective algorithms that could be used to provide model-reasoning.  This provides analysts and stakeholders a way to identify critical gaps in the existing set of approaches.

The unpredictable brittleness of black-box algorithms has provoked concern and increased scrutiny. But black-box algorithms have also had extraordinary success in some settings. We are concerned that a desire to use these powerful algorithms -- combined with the facile strength of outcome-reasoning -- has led to overconfidence from non-technical stakeholders. In contrast, model-reasoning can be more technically challenging to understand and has put limitations on the algorithms that can be deployed. We think these limitations are important to recognize. It has been hard to discuss these limitations because, for any given black-box algorithm, understanding why it might fail or when it might fail is challenging. In contrast, understanding which settings are appropriate for black-box deployment only requires understanding how they are developed -- that is, using the Common Task Framework (CTF). The MARA(s) framework extends the CTF into real-world settings, by isolating four problem-features -- measurement, adaptability, resilience, and agnosis (``MARA'') -- that mark a problem as being more or less suitable for black-box algorithms. Further, we suggest that the compact notation MARA(s) makes it clear how the assessment of the problem features is a function of the stakeholders. We hope MARA(s) will help the two cultures of statistical modeling -- and our stakeholders -- communicate and reason about algorithms.

\section{Acknowledgments}
We are grateful to many for helpful feedback on early versions of this work, including Angel Christin, Mark Cullen, Devin Curry, David Donoho, Jamie Doyle, Steve Goodman, Karen Kafadar, Robert E. Kass, Johannes Lenhard, Mark Liberman, Joshua Loftus, Kristian Lum, Ben Marafino, Blake McShane, Art Owens and his lab, Cynthia Rudin, Kris Sankaran, Joao Sedoc, Dylan Small, Larry Wasserman, Wen Zhou, The RAND Statistics Group, The Harvard Institute for Quantitative Social Science, The Quantitative Collaborative at the University of Virginia, and The Human and Machine Intelligence Group at the University of Virginia.  Finally, we thank the anonymous reviewers whose comments improved this paper.

\bibliography{bib}

\end{document}